\documentstyle[aps,prl,graphicx]{revtex}

\begin{document}

\title{Percolation line of
stable clusters in supercritical fluids}

\author{X. Campi, H. Krivine, N. Sator }
 
\address{Laboratoire de Physique Th\'eorique et Mod\`eles Statistiques
\footnote{Unit\'e de recherche de l'Universit\'e de Paris XI associ\'ee au
CNRS \\ campi@ipno.in2p3.fr} \\ Universit\'e Paris-Sud, B\^at. 100 \\ 91405
Orsay Cedex, France} \maketitle
\vspace*{.5cm}

\newcommand{\be}{\begin{equation}}
\newcommand{\ee}{\end{equation}}

\begin{abstract}
We predict that self-bound clusters of particles exist in the supercritical
phase of simple fluids. These clusters, whose internal temperature is lower
than the global temperature of the system, define a percolation line that
starts at the critical point. This line should be physically observable.
Possible experiments showing the validity of these predictions are discussed.

\vspace*{.5cm}
\end{abstract}

\twocolumn

Recent studies \cite{CHE99} of the supercritical phase of fluids reveal new
and unexpected phenomena that concern a variety of domains, ranging from
fundamental problems in cluster physics \cite{JIA92} to industrial
applications \cite{YOS97}. A question relevant to many of these domains is the
possible existence of clusters of particles. We predict in this Letter that
self-bound clusters of particles exist in the supercritical phase of simple
fluids and that these clusters define a physically observable percolation line
that starts at the critical point.

Clusters of particles in the supercritical phase have been considered in the
past, mainly from a theoretical point of view \cite{CoK80,KER89}. However,
these clusters are mathematical objects, introduced to match the percolation
threshold with the thermodynamical critical point. Furthermore, any critical
percolation line defined by these clusters has been considered as unphysical
and efforts have been made to eliminate them by a proper re-definition of the
clusters \cite{SW90,STA94}.

The results presented here are based on extensive numerical calculations
\cite{SAT00} of the cluster size distributions obtained from canonical Monte
Carlo and microcanonical Molecular Dynamics simulations of systems of $N \le
11664$ particles which interact through the Lennard-Jones potential:
$V(r_{ij})= 4\epsilon [(r_0/r_{ij})^{12}-(r_0/r_{ij})^6]$. The cutoff distance
of the potential is fixed at $3r_0$ and particles are confined in a cubic box
with periodic boundary conditions. The phase diagram of this fluid is
represented in figure 1 \cite{H-V,LJ_phase}. This phase diagram as well as the
results we will show below are insensitive to this particular choice of the
potential and can be considered generic for simple fluids. 

We assume that the system is at equilibrium when mean potential (kinetic)
energy reaches a stable value in the canonical (microcanonical) ensemble. In
the latter we take $2/3$ of the mean kinetic energy as a measure of the
temperature of the system.

\begin{figure}[h]
\begin{center}
\includegraphics[angle=-90,scale=0.4]{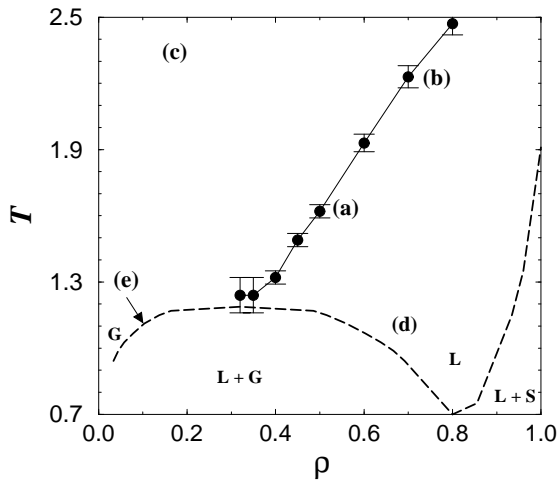}
\caption{\it \label{f:fig1} Phase diagram of the Lennard-Jones
fluid. The dashed line indicates the boundaries of the gas (G), liquid
(L) and solid phases (S) \protect\cite{H-V,LJ_phase}. The continuous
line (main purpose of this Letter) is the critical percolation line
calculated with the condition of stability of clusters. Temperature
$T$ and density $\rho$ are in units of $\epsilon$ and $r_{0}^{-3}$
respectively.}
\end{center}
\end{figure}

The cluster size distributions are calculated once thermal equilibrium is
reached. Physical clusters are defined according to a prescription proposed by
Hill \cite{HIL53}: At a given time, two particles are linked if their
potential energy exceeds their relative kinetic energy. A set of linked
particles forms a cluster. We have checked that using this procedure, for
equilibrium configurations, most clusters are stable by particle emission,
namely the kinetic energy of each particle relative to the center of mass of
the cluster is less than the sum of the potential energies due to the other
particles of the cluster.
\begin{figure}
\includegraphics[angle=-90,scale=0.6]{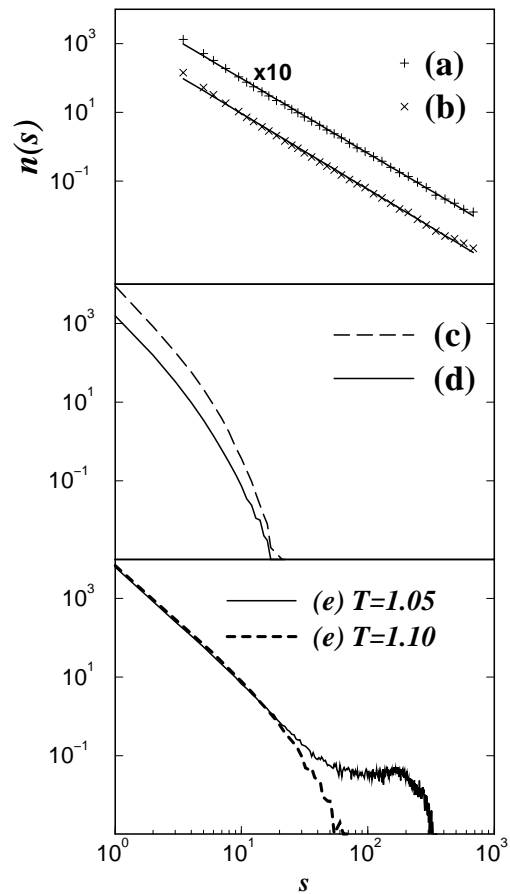}
\vspace{.5cm}

\caption{\it \label{f:fig2} Cluster size distributions $n(s)$ at points
(a)-(b), (c)-(d) and (e) of the phase diagram of figure 1. For curve (d), the
contribution of the largest cluster $S_{max}$ is sharply peaked beyond the
boundaries of the figure.}
\end{figure}

The mean cluster size distributions $n(s)$ ($s=1,2\cdots,N$) calculated at the
points $(a,b,c,d,e)$ of the phase diagram are displayed in figure 2. One
remarks that at points $(a)$ and $(b)$ the distributions can be fitted (for
$s>3$) by the same power law distribution $n(s) \sim s^{-\tau}$ with $\tau=
2.20 \pm 0.05$. In fact, the same power law behavior is found along the full
line of figure 1. This line $T_{p}(\rho)$ is defined as the locus of points
where $m_2$ reaches its maximum value \cite{CAM88}. The $m_{k}$ are the usual
moments of $n'(s)$ \cite{STA94}, where $n'(s)$ is the distribution of finite
size clusters, excluding the largest one. We find that to the right of this
line there exists a percolating cluster. Its size $S_{max}/N$ vanishes as
$(T_{p}(\rho)-T)^\beta$. In the vicinity of that line the moments behave as
$m_{2} \sim \vert T-T_{p}(\rho)\vert^{-\gamma}$ and $m_{3}/m_{2} \sim \vert
T-T_{p}(\rho)\vert^{-1/\sigma}$. The values we find for these exponents are in
agreement with those of random percolation theory (see table 1.) A more
rigorous determination of the ratios of critical exponents can be made using
finite size scaling relations. As a function of the size $N$ of the system,
one expects just on the critical line the behaviors: $S_{max}\sim
N^{1-\beta/3\nu}$, $m_2 \sim N^{\gamma/3\nu}$, $m_3/m_2 \sim
N^{1/3\sigma\nu}$, $<S_{max}^2>-<S_{max}>^2 \sim N^{1+\gamma/3\nu}$
\cite{STA80}, where $\nu$ is the critical exponent associated to the two-point
correlation function \cite{STA94}. These quantities, corresponding to point
$(b)$ of the phase diagram, are displayed as a function of $N$ in figure 3.
One observes that beyond $N \approx 50$ the expected power law behavior is
very well satisfied. The values we find for the ratios of critical exponents
are also in good agreement with those of random percolation (table 1.). We get
the same agreement at other points of the line. These results strongly suggest
that {\it clusters of self-bound sets of particles define a critical
percolation line (CPL), characterized by the universal exponents of random
percolation}. In addition, the present numerical calculations show that,
within the uncertainties inherent to the finite size of the system and to the
critical slowing down, this percolation curve ends at the (thermodynamical)
critical point $(T_{c},\rho_{c})$. Just at this point, for the same technical
reasons, we are unable to calculate accurately the corresponding critical
exponents. Extrapolating the exact results of Kasteleyn-Fortuin \cite{KF69}
and Coniglio-Klein for the lattice gas model \cite{CoK80}, one expects to get
at the (thermodynamical) critical point (and only at this point) the critical
exponents of the Ising model.

\begin{table}
\begin{center}
\begin{tabular}{|c|c|c|} 
           & Percolation &Present Work    \\ \hline
$\beta $    & 0.41 &  $0.4\pm .1$\\ \hline 
$\gamma $   & 1.80  & $1.6\pm .3$  \\ \hline 
$\sigma $   & 0.45  & $0.44\pm .05$ \\ \hline    
$\tau $     & 2.18  & $2.20\pm .05$ \\ \hline \hline
$\beta/\nu $ & 0.47  & $0.48\pm .02 $ \\ \hline
$\gamma/\nu $ & 2.05 & $ 2.0\pm .05 $ \\ \hline
$1/\sigma\nu $ & 2.53 & $ 2.60\pm .05 $ 

\end{tabular}
\caption{\it Critical exponents associated with the cluster size
distributions, in 3d Random Percolation \protect\cite{STA94} and
present work.}
\end{center}
\end{table}

 At $\rho < \rho_{c}$ or $T < T_{c}$ we do not see any signal of critical
(percolation) behavior. The distribution $n(s)$ decreases much faster than
$s^{-2.2}$ and clearly deviates from a power law form. However, a
``macroscopic'' cluster appears as soon as one penetrates into the two-phase
region. This is clearly seen in figure 2, where we plot two distributions
$n(s)$ around point $(e)$ of the phase diagram, with temperatures of $T=1.10$
and $T=1.05$, just above and below the liquid-gas coexistence curve. This
sharp signal corresponds very well to the crossing of the coexistence curve.
Indeed, small change in $T$ induces a drastic change in $n(s)$ for large $s$
\cite{rq1}.

We have checked that all the above findings do not depend on the specific
definition of stable clusters: Using two different definitions,
\cite{DOR93,PUE99}, both based on minimization procedures of the interaction
energy between different clusters, we get cluster distributions that are
almost identical \cite{rq2}.

It is interesting to emphasize that the CPL, characterized by the critical
exponents of random percolation, is found, on a deterministic dynamical
framework, {\it without} any explicit reference to a random (site or bond)
percolation mechanism. This is even more striking when using definitions
\cite{DOR93,PUE99}, for which clusters result from a global energy balance
(and not simply from a bond activation prescription).

\begin{figure}
\includegraphics[angle=-90,scale=0.4]{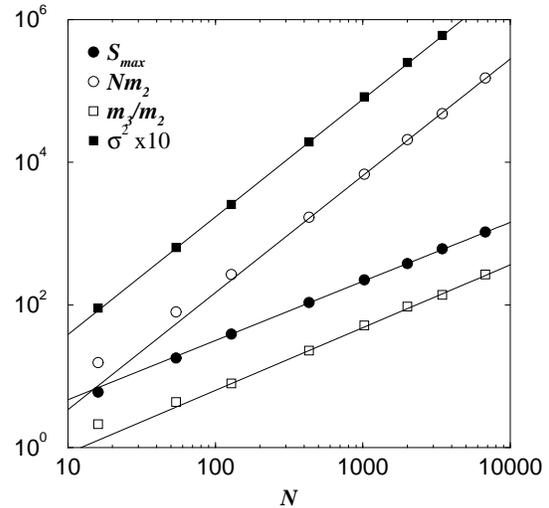}
\caption{\it \label{f:fig3} Finite size scaling behavior of $S_{max}$, $m_2$,
$m_3/m_2$ and $\sigma^2=<Smax^2>-<Smax>^2$ calculated at point (b) of the
phase diagram. The straight lines are the best fits in the range $50 < N <
11664$.}
\end{figure}

A this point, a question that arises naturally is the link between this CPL
and the critical percolation line of references \cite{CoK80,KER89}. We recall
that the latter is found in the lattice gas model when choosing the Coniglio
and Klein definition of clusters \cite{CoK80}. In ref.\cite{CKP99} it was
shown that this definition is (nearly) equivalent, in a lattice gas model, to
a condition of stability of clusters by monomer emission. In this sense, the
CPL can be seen as a generalization to a realistic fluid.

It is also interesting to recall that, when defining clusters with a real
space criteria, for instance as sets of particles that are two by two at a
distance less than some distance $d_{c}$, a percolation line also exists
\cite{YOS97,GEI82,HEY88}, but its position depends crucially on the choice of
$d_{c}$. However, this percolation line may be relevant in situations in which
physical clusters are sets of particles close in {\it{real}} space, for
example in conductivity experiments in which neighboring atoms exchange
electrons in overlapping orbits.

As a natural consequence of the definition of clusters based on a stability
condition, we find that the internal `` effective temperature'' $T_{eff}$ of
clusters \cite{rq3} is always less than the global temperature $T$ of the
system. Using the molecular dynamics calculation, we find that $T_{eff}$ grows
as a function of $s$ from zero for $s=1$, to a limiting value for $s\gtrsim
100$. In the supercritical phase, this limiting temperature, that depends
essentially on the density of the system, ranges between $T_{eff}=0.7$ and
$T_{eff}=1.0$, i.e. small clusters behave as ``solid-like crystals'' while the
large ones behave as ``liquid-like droplets'', as can be seen considering the
complete phase diagram. We have also observed that along the CPL the total
energy of the system remains almost constant. The origin of this energy
invariance, that results from a subtle balance between internal, center of
mass kinetic energies and inter-cluster potential energies, is not understood.

Finally, we briefly discuss the possibilities to observe experimentally the
CPL. Notice first that this line should not be confused with the Fisher-Widom
line \cite{FIS69}, or with the extrapolation of the rectilinear diameters line
\cite{NIM00}. Those two lines divide the supercritical region into a gas like
and a liquid like domain with probably no relevance on clustering.

Many experiments dealing with critical behavior of binary fluids (which belong
to the same universality class as liquid-gas and $3d$ Ising model) have been
performed \cite{STA72}. Observations of concentration fluctuations in the
mixture of isobutyric acid and water have been done in the vicinity of the
critical point \cite{GUE89} and a fractal dimension $D_f=2.8\pm0.1$ of
clusters has been determined. The value we determine along the CPL is
$D_f=1/\sigma\nu=2.60 \pm 0.05$. However, the relationship between these
clusters, defined from persistent density fluctuations and the stable clusters
is not yet clear. Work is in progress in this direction. The mobility of
$H^{+}$ ions in the same binary mixture has also been studied very recently
\cite{BON00}. The sharp decrease of the ion mobility observed for $T>T_{c}$ as
the critical concentration is exceeded, has been associated to the appearance
of a percolation line of dynamical clusters. The position of this curve in the
$\rho-T$ plane would suggest however a closer connection with clustering in
real space \cite{HEY88}.

In a different domain, early experiments on the effusion of a fluid through a
pin hole \cite{MIL56} have shown the presence of stable clusters (dimers and
trimers) in the gas phase. In these experiments, the mass yield of escaped
clusters with a given velocity can be related to the moments $m_{k}$ of the
cluster size distribution $n'(s)$. Unfortunately, the interesting regime,
namely the crossing of the CPL has not yet been explored. It should also be
possible to isolate stable clusters by a sudden disassembly of a piece of
fluid into droplets, like in the fragmentation of atomic nuclei or atomic
aggregates \cite{NUCLEI,DROPS}. Several experiments \cite{CAM88,NUCLEI} show
indeed, as a function of the excitation energy, an evolution of the fragment
size distributions $n(s)$ that suggests a crossing of the CPL. However, due to
the small size of the system, the results are still inconclusive.

In summary: Large scale and long time Monte Carlo and Molecular Dynamics
calculations suggest that energetically stable clusters are present in a
supercritical simple fluid. These clusters, that are cooler than the system,
percolate below a critical line that ends at the (thermodynamical) critical
point. Various experiments, within the reach of present day experimental
techniques, are suggested to show the presence of these stable clusters and
the existence of this critical percolation line.

We have benefited from fruitful discussions with O. Martin and E. Plagnol. O.
Bohigas is specially acknowledged for his critical remarks.


\begin{thebibliography}{99}
\bibitem{rq2} This is strictly true for the finite size clusters. However, the
size of the percolating cluster (if it exists) may depend on the cluster
definition.
\bibitem{rq3} It would be more appropriate to talk about mean kinetic energy,
because the distribution of relative velocities deviates from a Maxwellian,
due to the definition of clusters which reduces the high velocity tail.
\end{thebibliography}
\end{document}